\documentclass[%
 reprint,
 amsmath,amssymb,
 aps,
]{revtex4-1}

\usepackage{graphicx}
\usepackage{dcolumn}
\usepackage{bm}

\begin{document}

\preprint{APS/123-QED}

\title{Quantum Ho\v{r}ava-Lifshitz cosmology and the quantum nature of coupling corrections of HL gravity }

\author{Laysa G. Martins} \email{laysamartinsymail@yahoo.com.br}
\author{Jos\'{e} A. C. Nogales}%
\email{jnogales@dfi.ufla.br}
\affiliation{Departamento de F\'{i}sica (DFI) and Museu de Historia Natural (MHN), Universidade Federal de Lavras (UFLA),Lavras-MG, Caixa postal 3037, CEP 37000-000, Brazil.
}%




\date{\today}

\begin{abstract}
In this work were studied quantum models of a Friedmann-Robertson-Walker (FRW) cosmology
in the framework of the gravity's theory proposed by Ho\v{r}ava, the so-called Ho\v{r}ava-Lifshitz theory
of the gravity. It was used the Ho\v{r}ava theory for the projectable Ho\v{r}ava-Lifshitz (HL) gravity without
the detailed balance condition. Following the quantization of the model in the context of 
Wheeler-DeWitt   approach and  taking in account the ordering factor for operators  were found the cosmological wave function. Solutions were studied and the results were discussed for some
particular cases close of initial singularity. The resulting wave functions were used to investigate
the possibility of to avoid the classical singularities due to quantum effects and for analyzing
the entanglement entropy. In the ultraviolet phase were found the existence of cosmological
wave function with a relation between the ordering factor and coupling constants showing their quantum nature,  then it was possible
to provide an explicit evolution of the cosmological entanglement entropy in this stage. The
interpretation of Bohm-De Broglie was used to discussion of the solutions.
\end{abstract}


\maketitle
%

\section{\label{sec:level1}Introduction}
\qquad It has been questioned for some time ago about the fact of general relativity not be a fundamental theory of gravity and the need to be changed when considering high energies. One proposal, among many, that complements Einstein's theory at high energies, was presented in 2009 by Ho\v{r}ava. This  theory of gravity proposes sensitive modifications  in general relativity when we are working in the ultraviolet range (UV), since the standard of quantization procedure of general relativity presents difficulties related with nonrenormalizability, the proposal of Ho\v{r}ava seems to remove this anomaly.\\ 
\indent After the publication of Ho\v{r}ava theory \cite{Horava}, several articles has been published in this area, which led to significant results in quantum gravity and cosmology. The proposed Ho\v{r}ava has been inspired by a phase transition studied in solid state physics, in particular the work of Lifshitz. So this theory is known in the literature as theory Ho\v{r}ava-Lifshitz.\\
\indent The model Horava-Lifshitz far to propose modifications in Einstein's gravitation UV regimen makes this renormalizable and recover general relativity standard for IR regimens limit. For this to happen it is necessary to enter higher order derivatives in the Lagrangian, establishing the renormalization. However this insertion results in some problems, such as higher order derivatives in time, which, when applied in theory lead to phantom fields \cite{Sotiriou2009a}. \\
\indent Ho\v{r}ava to realize that space higher order derivatives contribute to the renormalization theory, while the time derivatives of higher order produce ghosts, he had the idea to construct a theory that used only the spatial derivatives of higher order, for it used the work of Lifshitz in physics of solid state \cite{Garattini} thus proposes a re-scaling anisotropic in space-time in the UV regime.\\
\indent To enter the partial derivatives of higher order in the Lagrangian without the necessary enter such time derivatives, an interesting possibility is to use the formalism developed by Richard Arnowitt, Stanley Deser and Charles W. Misner (ADM) that was developed in the late 1950s \cite{Gourgoulhon}. The ADM formalism is a possibility of approach to quantum gravity, through the Hamiltonian formulation of general relativity \cite{Gourgoulhon}. This formalism consists in foliate the space-time, getting three-dimensional spatial hypersurfaces that evolve over time.  \\
\indent Then applying the Dirac quantization procedure, from the quantum prescriptions is obtained equation analogous to the Schrodinger equation, known as the Wheeler-DeWitt equation of \cite{DeWitt}. This equation provides the dynamics of the quantum system and determines the wave function of the universe. \\
\indent The Hamiltonian describing this system has restrictions. The resolution of the Wheeler-DeWitt equation for the case where the space-time is not homogeneous, it is very difficult. But if the space-time has certain symmetries, the Wheeler-DeWitt equation can be reduced to a simple equation and may be solved using a finite number of degrees of freedom. This is called approximation of mini-superspace, which will be considered in this work. \\
\indent Beginning with the ADM representation of the action corresponding to this model, we construct the Lagrangian in terms of the minisuperspace  variables without matter \cite{Bertolami} and show that in comparison with the usual Einstein-Hilbert gravity, there are some correction terms coming from the Ho\v rava theory for the projectable Ho\v rava-Lifshitz (HL) gravity without the detailed balance condition in the very early universe (see for example \cite{wang1} and \cite{wang2}) . In addition  to perform the Wheeler- De Witt equation we consider  the ordering factor in the ultraviolet stage since it will be relevant in this stage and negligible at classical transition \cite{Hawking} \cite{Kontoleon}.\\
\indent The growth of the entanglement entropy with the scale factor provides a new statistical notion of arrow of time in quantum gravity \cite{Chen}. The growth of entanglement in the ultraviolet phase in HL gravity will provides a mechanism for the production of the quantum correlations present at the beginning of universe we test the growth  of entanglement entropy for all cases considered in this work. \\
\indent In this work we take, from several interpretations of quantum mechanics \cite{Pinto,PintoNeto}, the interpretation of de Broglie-Bohm \cite{Durr} , because this context not need a classic observer. \\
\indent This paper is organized as follows: in  the second section we resume the cosmological toy model of Ho\v{r}ava-Lifshitz. In Sections III A we present the factor order of momentum operator, which will be used to quantize the Hamiltonian system. Then the topic III B considers the case in which this very close to the singularity, with this the Hamiltonian can be rewritten considering only the some terms of the potential that are dominant than other terms and calculate explicitly the entanglement entropy. In the following subsections, we studied regions very close to the singularity as described in Sections III B, but that are also relevant. To this is added more terms to the potential considered in section III B. In subsections III C and III D this study was done to the spherical and hyperbolic geometries respectively. It is determined the quantum effects on the system and were analyzed the contribution of  factor order in the solutions and the explicit evolution of entanglement entropy.

\section{\label{sec:level2}Cosmological toy model for Ho\v rava's proposal}
\qquad The action for the projectable HL gravity without detailed balance is given by Sotiriou, Visser and Weinfurtner \cite{Sotiriou2009b}:
\begin{eqnarray}
 S &=& \frac{M_{pl}^{2}}{2}\int_{\mathcal{M}}d^{3}xdtN\sqrt{h}[K_{ij}K^{ij}-\lambda K^{2}-g_{0}M_{pl}^{-2}-\nonumber \\
 &-& g_{1}R-M_{pl}^{-2}(g_{2}R^{2}+ g_{3}R_{ij}R^{ij}) - M_{pl}^{4}(g_{4}R^{3}+\nonumber \\ &+&g_{5}RR^{i}_{j}R^{j}_{i}+g_{6}R^{i}_{j}R^{j}_{k}R^{k}_{i}+g_{7}R\nabla^{2}R+\nonumber \\
 &+& g_{8}\nabla_{i}R_{jk}\nabla^{i}R^{jk})], \label{C1}
\end{eqnarray}
\noindent where $M_{pl}=(8\pi G)^{-1/2}$ is the Planck mass, $K_{ij}$ are the components of extrinsic curvature, $h$ is the determinant of $h_{ij}$, $R_{ij}$ are the Ricci components of the 3-metric, $R$ is the scalar Ricci for $h_{ij}$ and $K$  the trace of $K_{ij}$. The constants  $\lambda$ and $g_{n}=(0,1,...,8)$ are coupling constants. As shown by Vakili e Kord (2013), when 
$\Lambda=g_{0}M_{pl}^{2}/2$, $g_{1}=-1$ e $\lambda=1$, the RG is recovered a RG in infrared limit.\\
\indent In this work we analize the Lema\^{i}tre-Friedmann-Robertson-Walker (LFRW) in the Ho\v{r}ava framework given by:
\begin{eqnarray}
 ds^{2}=-N^{2}(t)dt^{2}+a^{2}(t)\bigg[\frac{dr^{2}}{1-kr^{2}}+r^{2}(d\vartheta^{2}+\sin^{2}\vartheta d\varphi^{2})\bigg], \label{tres.0}
\end{eqnarray}
where $N(t)$ is the lapse function, $a(t)$ is the scale factor and  $k$  is a constant representing the curvature of the space. We rewrite this metric as follows:
\begin{eqnarray}
 ds^{2}=-N^{2}(t)dt^{2}+h_{ij}dx^{a}dx^{b}, \nonumber
\end{eqnarray}
\noindent where 
\begin{eqnarray}
 h_{ij}=a^{2}(t)\textrm{diag}\left(\frac{1}{1-kr^{2}},r^{2},r^{2}\sin^{2}\vartheta\right), \nonumber
\end{eqnarray}
\noindent here $h_{ij}$  is the metric of the three-dimensional slices. \\
\indent Rewriting the equation (\ref{C1}) in terms of $\Lambda=g_{0}M_{pl}^{2}/2$ and $g_{1}=-1$, we obtain :
\begin{eqnarray}
 S &=& \frac{M_{pl}^{2}}{2}\int_{\mathcal{M}}d^{3}xdtN\sqrt{h}[K_{ij}K^{ij}-\lambda K^{2}+R-2\Lambda-\nonumber \\
 &-&M_{pl}^{-2}(g_{2}R^{2}+g_{3}R_{ij}R^{ij})- M_{pl}^{4}(g_{4}R^{3}+g_{5}RR^{i}_{j}R^{j}_{i}+\nonumber \\
 &+&g_{6}R^{i}_{j}R^{j}_{k}R^{k}_{i}+g_{7}R\nabla^{2}R+g_{8}\nabla_{i}R_{jk}\nabla^{i}R^{jk})], \label{C2}
\end{eqnarray}
\noindent Furthermore, the requirement that this action be \hyphenation{e-qui-va-lent}equivalent to the standard Einstein-Hilbert action in the IR limit requires that the running constant $\lambda$ takes its relativistic value $\lambda = 1$.\\
\indent The extrinsic curvature tensor, which measures how the spatial slices in the ADM decomposition of space-time curves with respect to external observers, is defined by:
\begin{eqnarray}
 K_{ij}=\frac{1}{2N} \left(\nabla_{j}N_{i}+\nabla_{i} N_{j} -\frac{\partial h_{ij}}{\partial t}\right), \nonumber
\end{eqnarray}
\noindent where $N_{i}$ is the shift vector and $\nabla_{j} N_{i}$ represents the covariant derivative with respect to $h_{ij}$, for metric  (\ref{tres.0}) we have:
\begin{eqnarray}
 K_{ij}K^{ij}=\frac{3\dot a^{2}}{N^{2}a^{2}} \qquad \textrm{e} \qquad K=-\frac{3\dot a}{Na}, \label{C3}
\end{eqnarray}
\noindent here $\dot a$ where a dot represents differentiation with respect to t. The Ricci tensor and the Ricci scalar correspond to the 3-geometry $h_{ab}$ can be obtained as:  
\begin{eqnarray}
 R_{ij}=\frac{2kh_{ij}}{a^{2}} \qquad \textrm{e} \qquad R=\frac{6k}{a^{2}}. \label{C4}
\end{eqnarray}
\noindent Substituting equations (\ref{C3}) and (\ref{C4}) in equation (\ref{C2}), the action is given by:
\begin{eqnarray}
 S &=& \frac{3V_{0}M_{pl}^{2}(3\lambda-1)}{2}\int dtN\bigg\{-\frac{a\dot a^{2}}{N^{2}}+\frac{6ka}{3(3\lambda-1)}- \nonumber \\
 &-& \frac{2\Lambda a^{3}}{3(3\lambda-1)}-M_{pl}^{-2}\left[\frac{12k^{2}(3g_{2}+g_{3})}{3a(3\lambda-1)}\right]-\nonumber \\ &-&M_{pl}^{-4}\left[\frac{24k(9g_{4}+3g_{5}+g_{6})}{3a^{3}(3\lambda-1)}\right]\bigg\}, \label{C5}
\end{eqnarray}
\noindent where $V_{0}=\int d^{3}x\sqrt{h}$ is the integral over spatial dimensions.\\
\indent If we fixed $ 3V_{0}M_{pl}^{2}(3\lambda-1)/2=1 $, we can write the Lagrangian like :
\begin{equation}
L=\frac{N}{2}\bigg(-\frac{a\dot a^2}{N^2}+g_{c}ka-g_{\lambda}a^{3}-\frac{g_{r}k^{2}}{a}-\frac{g_{s}k}{a^{3}}\bigg), \label{C9}
\end{equation}
\noindent where the coefficients $g_{i}$ are defined in Sotiriou, Visser e Weinfurtner \cite{Sotiriou2009b}, like:
\begin{eqnarray}
g_{c}&=&\frac{2}{3\lambda-1},\quad g_{\lambda}=\frac{2\Lambda}{3(3\lambda-1)},
\quad g_{r}=6V_{0}(3g_{2}+g_{3}), \nonumber \\ \quad g_{s}&=&18V_{0}^{2}(3\lambda-1)(9g_{4}+3g_{5}+g_{6}),
\end{eqnarray}
\noindent where the dimensionless coupling constants $g_{c}>0$ are related with the curvature coupling constant, $g_{\lambda}$ is related with the cosmological constant, $g_{r}$ with o behaviour of radiation and  $g_{s}$ like stiff matter, all of this in correspondence of equation of state ($p=\rho$). The coupling constants $g_{r}$ and $g_{s}$  can be either positive or negative as their signal does not alter the stability of the HL gravity \cite{Maeda}. \\
\indent The Hamiltonian for this model can be obtained from its standard definition:
\begin{equation}
H =\dot a\Pi_{a}-L, \nonumber 
\end{equation}
where the canonical momentum is defined by : 
\begin{equation}
\Pi_{a}=\frac{\partial L}{\partial \dot a}=-\frac{a\dot a}{N}, \nonumber
\end{equation}
\noindent then the Hamiltonian is given by: 
\begin{eqnarray}
H &=& \dot a\bigg(-\frac{a\dot a}{N}\bigg)-\frac{N}{2}\bigg(-\frac{a\dot a^{2}}{N^{2}}+g_{c}ka-g_{\lambda}a^{3}-\nonumber \\
&-& \frac{g_{r}k^{2}}{a}-\frac{g_{s}k}{a^{3}}\bigg),
\nonumber \\
H &=& -\frac{1}{2}\bigg(\frac{N}{a}\bigg)\bigg[-\Pi_{a}^{2}-g_{c}ka^{2}+g_{\lambda}a^{4}+g_{r}k^{2}+\frac{g_{s}k}{a^{2}}\bigg), \label{C6}
\label{5}
\end{eqnarray}
\noindent This Hamiltonian we do not have matter terms only geometric terms.

\section{\label{sec:level2} Solutions of the Wheeler-DeWitt equation }

\subsection{Factor ordering}

\indent In this section we are show a factor ordering  for the Hamiltonian constraint of cosmological toy models. The  factor  ordering  problem  in  cosmology  was  studied  by  many  researchers. Louko  and Barvinsky \cite{Louko,Barvinsky} focused in the D'Alembertian ordering, i.e.  the covariant ordering.  The non-dynamical states was considered by Spiegel \cite{Spiegel}. The influence of the ordering on solutions and interpretation of wave function was also studied by Kontoleon \cite{Kontoleon}.  The factor ordering problem was also studied in full quantum gravity by DeWitt and Kucha\v{r} \cite{DeWitt,KUCHAR}.\\
\indent   The aim of section  is the formulation of these issue for models defined in Ho\v{r}ava mini-superspaces given in the last section. In order to obtain the quantum equation for (\ref{C6}), we use the proposal given by \cite{Hawking,Nelson}:
\begin{equation}
\Pi_{a}^{2}=\frac{1}{a^{p}}\frac{\partial}{\partial a}\left(a^{p}\frac{\partial}{\partial a}\right), \label{C7}
\end{equation} 
\noindent  where $p$ indicates the ambiguity of the ordering of factors $a$ and $\Pi_{a}$. Thus substituting the eq.(\ref{C7}) in eq.(\ref{C6}),  we obtain:
\begin{equation}
\left(\frac{\partial^2}{\partial a ^2}+\frac{p}{a}\frac{\partial}{\partial
a}+g_{c}ka^{2}-g_{\lambda}a^{4}-g_{r}k^{2}-\frac{g_{s}k}{a^{2}}\right)\psi(a)=0. \label{C10}
\end{equation}

\noindent This equation is the  Wheeler-DeWitt equation, where $\psi=\psi(a)$ is the quantum wave function for our universe. In this work we study the solution for this equation considering $a\ll1$, solutions for other condition was analyzed by for example by Bertolami \cite{Bertolami}.

\subsection{\label{sec:level2}General solution for $a(t)\sim 0$}
\qquad In this section we present solution that represent the classical singularity in cosmological model. We consider the factor scale   $a\rightarrow0$ in the equation  (\ref{C10}). So the we easily identify the predominant terms in the potential for Hamiltonian in this equation given by:  
\begin{equation}
V(a)=-\frac{g_{s}k}{a^{2}}, \nonumber
\end{equation}
\noindent 
so the equation (\ref{C10}) becomes :
\begin{equation}
\left(\frac{\partial^2}{\partial a ^2}+\frac{p}{a}\frac{\partial}{\partial a}-\frac{g_{s}k}{a^{2}}\right)\psi(a)=0. \label{C11}
\end{equation}
\indent  The cases in which $a\gg1$ have been studied and show that they decay in the standard cosmology \cite{Bertolami,Vakili}. 
For the wavefunction $\psi(a)$ 
the ideas will be used are of De Broglie-Bohm, whose form $\psi(a)$ 
It is given by:
\begin{equation}
\psi(a)=R(a)e^{iS(a)}, \label{C12}
\end{equation} 
\noindent where $R=R(a)$ is the real part and  $S=S(a)$ is the phase of the wave function. From De Broglie-Bohm theory, see for example Pinto-Neto \cite{PintoNeto}, can be established
the probability density and velocity, respectively given by:
\begin{equation}
\rho=R^2 \quad \textrm{e} \quad \vec{v}=\vec{\nabla} S. \label{C13}
\end{equation}
\noindent Substituting the equation  (\ref{C12}) in the equation  (\ref{C11}), we obtain:
\begin{equation}
\left(\frac{\partial^2}{\partial a ^2}+\frac{p}{a}\frac{\partial}{\partial
a}-\frac{g_{s}k}{a^{2}}\right)R(a)e^{iS(a)}=0. \nonumber
\end{equation}
\noindent From the last equation we obtain, after separate the imaginary and real part, the following two equations: 
\begin{eqnarray}
(S'(a))^2+Q(a)+V(a)&=&0, \label{C14}\\
S''(a)+\frac{2R'(a)S'(a)}{R(a)}+\frac{p}{a}S'(a)&=&0, \label{C15}
\end{eqnarray}
\noindent where the equation (\ref{C14}) gives the dynamics of the system (equation Hamilton-Jacobi), where the second term of this equation is knowing like quantum potential, and is defined by:
\begin{equation}
Q(a)=-\left(\frac{R''(a)}{R(a)}+\frac{p}{a}\frac{R'(a)}{R(a)}\right). \nonumber
\end{equation}
\noindent In quantum gravity, when the scale factor is small, quantum effects must be considered,  conversely we can use the semi classically approach developed by  Wentzel-Kramers-Brillouin (WKB), in this case  the quantum potential is not considered because is negligible,  however, there is a correlation between classical and quantum solutions given by  $\partial S/\partial a$. So when the scale factor is small $a\rightarrow0$, it should be considered the quantum effects, thus the WKB approximation is not well founded.\\
\indent The equation  (\ref{C15}) resemble the continuity equation. For the model considered in this article, this equation is given by:
\begin{equation}
j^{a}=\frac{i}{2}a^{p}(\psi^{*}(a)\partial _{a}\psi(a)-\psi(a)\partial_{a}\psi^{*}(a)), \nonumber
\end{equation}
\noindent where $\psi^{*}(a)=R(a)e^{-iS(a)}$, is the complex conjugate of the function (\ref{C12}) and $a^{p}$ is the term related with the factor ordering. From this equation we find that:
\begin{equation}
j^{a}=-a^{p}[(R(a))^{2}S'(a)]. \nonumber
\end{equation}
\indent The solution of equation (\ref{C11}) which describe the classical singularities in the beginning of the universe is given by:
\begin{eqnarray}
\psi(a)&=&c_{1}a^{-\frac{1}{2}(p-1)+\frac{1}{2}\sqrt{(p-1)^{2}+4kg_{s}}}+\nonumber \\
&+& c'_{2}a^{-\frac{1}{2}(p-1)-\frac{1}{2}\sqrt{(p-1)^{2}+4kg_{s}}}, \label{C16}
\end{eqnarray}
\noindent where $c_{1}$ e $c'_{2}$ are constants. Assume that the constant $c_{1}$ is a real number and the constant $c'_{2}$, a complex quantity, so the above expression can be rewritten as:
\begin{eqnarray}
\psi(a)&=&c_{1}a^{-\frac{1}{2}(p-1)+\frac{1}{2}\sqrt{(p-1)^{2}+4kg_{s}}}+\nonumber \\
&+& ic_{2}a^{-\frac{1}{2}(p-1)-\frac{1}{2}\sqrt{(p-1)^{2}+4kg_{s}}}, \label{funcaoonda}
\end{eqnarray}
\noindent note that the wave function described by (\ref{funcaoonda}) is of the form $\psi(a)=U+iW=R(a)e^{iS(a)}$, so we can express the real part and imaginary given respectively by:
\begin{eqnarray}
R^{2}=U^{2}+W^{2}, \qquad S=\tan^{-1}\bigg(\frac{W}{U}\bigg),
\end{eqnarray}
\noindent so it follows that:
\begin{eqnarray}
R &=& \bigg[\left(c_{1}a^{-\frac{1}{2}(p-1)+\frac{1}{2}\sqrt{(p-1)^{2}+4kg_{s}}}\right)^{2}+ \nonumber\\
&+& \left(c_{2}a^{-\frac{1}{2}(p-1)-\frac{1}{2}\sqrt{(p-1)^{2}+4kg_{s}}}\right)^{2}\bigg]^{\frac{1}{2}}, \label{real} \\
S &=& \tan^{-1}\left(\frac{c_{2}a^{-\sqrt{(p-1)^{2}+4kg_{s}}}}{c_{1}}\right),
\end{eqnarray}
\noindent as is the case analyzing the $a\ll1$ can be made to approach $\tan^{-1}\theta\approx\theta$, so the phase $S$ is given by:
\begin{eqnarray}
S &=&\left(\frac{c_{2}a^{-\sqrt{(p-1)^{2}+4kg_{s}}}}{c_{1}}\right).
\end{eqnarray}
\indent  The bhomian trajectories are obtained by:
\begin{equation}
\dot a=-\frac{1}{a}\frac{\partial S}{\partial a}, \label{C22}
\end{equation}
\noindent this mean:
\begin{equation}
\dot a=\frac{c_{2}}{c_{1}}[(p-1)^{2}+4kg_{s}]^{\frac{1}{2}}a^{-\sqrt{(p-1)^{2}+4kg_{s}}-2}. \nonumber
\end{equation}
\noindent integrating the equation above,  we have:
\begin{equation}
a(t)=\left[A(t+t_{0})\right]^{\frac{1}{\sqrt{(p-1)^{2}+4kg_{s}}+3}}, \label{C999}
\end{equation}
\noindent where $A=(c_{2}/c_{1})[(p-1)^{2}+4kg_{s}+3\sqrt{(p-1)^2+4kg_{s}}]$. Analyzing the expression (\ref{C999}) have that $p\neq1\pm\sqrt{9+4kg_{s}}$ not to have uncertainty in the exponent and also that $(p-1)^2+4kg_{s}>0$ for the coefficient $A$ is real. So we have the situation, when $kg_{s}<0$, i.e. : $k>0$ and $g_{s}<0$ or $k<0$ and $g_{s}>0$. In the first case  In the first case, when $k >  0$, will have a spherical universe and from the constant $ g_ {s} $ follows that:
\begin{eqnarray}
g_{s}<0, \nonumber \\
288\pi(3\lambda-1)(9g_{4}+3g_{5}+g_{6})<0, \nonumber \\
\lambda<\frac{1}{3},
\end{eqnarray}
\noindent where $9g_{4}+3g_{5}+g_{6}>0$, so clearly this solution not decay in the RG  since $\lambda \not \to 1$ required by Ho\v{r}ava proposal. In the second case  $k<0$ and $g_{s}>0$, correspondent to hyperbolic universe, we have for  $g_{s}$ that:
\begin{eqnarray}
g_{s}>0, \nonumber \\
288\pi(3\lambda-1)(9g_{4}+3g_{5}+g_{6})>0, \nonumber \\
\lambda>\frac{1}{3},
\end{eqnarray}
we consider that  $9g_{4}+3g_{5}+g_{6}>0$, then RG is recovered at low energies.\\
\indent Therefore, it can be concluded that the solution (\ref{C999}) is a finite and regular function but valid for small \hyphenation{va-lues}values of the scale factor $a$, and $k$ and $g_ {s}$ should be mixed signals, as discussed above. With this solution it is possible to determine the probability density over time given by equation whose real part is given by (\ref{real}). Using the equations (\ref{real}) and (\ref{C999}), we get:
 \begin{equation}
 \rho(t)=\left\{ c_{1}^{2}[A(t+t_{0})]^{\frac{\alpha+\beta}{2\beta+3}}+c_{2}^2[A(t+t_{0})]^{\frac{\alpha+\beta}{\beta+3}}\right\}^{\frac{1}{2}}, \label{C998}
 \end{equation}
 \noindent where $\alpha=-(p-1)$ e $\beta=\sqrt{(p-1)^{2}+4kg_{s}})$. Note that the solution given by (\ref{C998}) shows that $\rho \rightarrow 0$, ie as $t\rightarrow 0$ the probability density decreases, meaning it is more likely to find the scale factor values to $t$ larger, or away from $t=0$ which gives rise to the singularity $a \to 0 $. Therefore, the singularity is removable.\\
\indent Following the ideas of Chen and Liu \cite{Chen}, who analyzed the behavior of entangled entropy for the cases in which the factor scale is very small, for the case above was possible to establish a relationship between the scale factor, entropy and order factor. For this, consider that entropy, according to Chen and Niu, can be written as:
\begin{eqnarray}
 \mathcal{S} = \int_{a_{0}}^{{a(t)}}\psi^{*}\psi\ln(\psi^{*}\psi)da + \mathcal{S}_{0}. \label{entropia}
\end{eqnarray}
\noindent To calculate the entropy in the case where the scale factor is very small, it will be used the wave function given by equation (\ref{funcaoonda}) and we will consider that $\alpha=0$ e $\beta=1/2$. Thus, we obtain that:
\begin{eqnarray}
 \mathcal{S} &=& 2\ln\left(c_{1}^{2}a(t)^{\frac{1}{2}}+c_{2}^{2}a(t)^{-\frac{1}{2}}\right)\left(\frac{1}{3}c_{1}^{2}a(t)^{\frac{3}{2}}+c_{2}^{2}a(t)^{\frac{1}{2}}\right)- \nonumber \\
 &-& \frac{2}{3}c_{1}^{2}\bigg[\frac{1}{3}a(t)^{\frac{3}{2}}-2\left(\frac{c_{2}}{c_{1}}\right)^{2}a(t)^{\frac{1}{2}}+ \nonumber\\ &+&2\left(\frac{c_{2}}{c_{1}}\right)^{3}\arctan\left(\left(\frac{c_{1}^{2}a(t)}{c_{2}^{2}}\right)^{\frac{1}{2}}\right)\bigg]- \nonumber \\
 &-& c_{2}^{2}\left[2a(t)^{\frac{1}{2}}-4\left(\frac{c_{2}}{c_{1}}\right)\arctan\left(\left(\frac{c_{1}^{2}a(t)}{c_{2}^{2}}\right)^{\frac{1}{2}}\right)\right] + \nonumber \\
 &+& \mathcal{S}_{0}, \label{entropia1}
\end{eqnarray}
\noindent where all the constants were included in $\mathcal{S}_{0}$. Equation (\ref{entropia1}) describes how entangled entropy varies in relation to the scale factor for the case in which has $p=1$. The numerical result is shown in Figure 1. It is observed that as the scale factor becomes smaller, the entropy decreases.

\begin{figure}[!htb]
\centering
\includegraphics[scale=0.5]{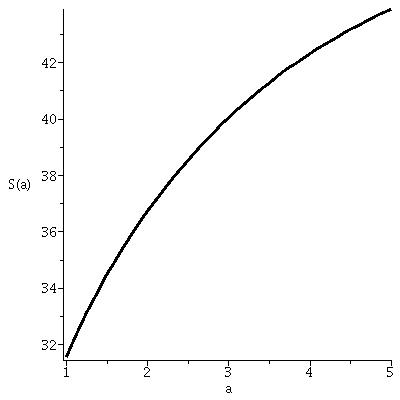}
\caption{Entangled entropy versus the scale factor, for the value of $p=1$.}
\end{figure}

\indent Can be established a relationship between entropy and the time, just replace the equation (\ref{C999}) in equation (\ref{entropia1}), so we have:
\begin{eqnarray}
\mathcal{S} &=& 2\ln\left(c_{1}^{2}[A(t+t_{0})]^{\frac{1}{2(\beta+3)}}+c_{2}^{2}[A(t+t_{0})]^{-\frac{1}{2(\beta+3)}}\right)* \nonumber \\
&*&\left(\frac{1}{3}c_{1}^{2}[A(t+t_{0})]^{\frac{3}{2(\beta+3)}}+c_{2}^{2}[A(t+t_{0})]^{\frac{1}{2(\beta+3)}}\right)- \nonumber \\
 &-& \frac{2}{3}c_{1}^{2}\bigg[\frac{1}{3}[A(t+t_{0})]^{\frac{3}{2(\beta+3)}}-2\left(\frac{c_{2}}{c_{1}}\right)^{2}[A(t+t_{0})]^{\frac{1}{2(\beta+3)}}+ \nonumber\\ 
 &+& 2\left(\frac{c_{2}}{c_{1}}\right)^{3}\arctan\left(\frac{c_{1}[A(t+t_{0})]^{\frac{1}{2(\beta+3)}}}{c_{2}}\right)\bigg]- \nonumber \\
 &-& 2c_{2}^{2}\bigg[[A(t+t_{0})]^{\frac{1}{2(\beta+3)}}- \nonumber \\
 &-&2\left(\frac{c_{2}}{c_{1}}\right)\arctan\left(\frac{c_{1}[A(t+t_{0})]^{\frac{1}{2(\beta+3)}}}{c_{2}}\right)\bigg]+\mathcal{S}_{0}. \label{entropia2}
\end{eqnarray}
\noindent Equation (\ref{entropia2}) describes the entropy function of time. Note that when $t\rightarrow0$ entropy is described in terms of a constant that characterizes the beginning of the universe.\\
\indent In the following two subsections two cases will be \hyphenation{a-na-ly-zed}analyzed close the singularity, taking in consideration \hyphenation{a-no-ther}another term potential, containing the constant $ g_ {r} $ in equation (\ref{C10}). This study will be done to the spherical and hyperbolic geometries respectively.

\subsection{\label{sec:level3}Closed universe}

\qquad The Wheeler-DeWitt equation (\ref{C10}) for the case that the universe is closed, becomes:
 \begin{eqnarray}
 \frac{d^{2}\psi(a)}{da^{2}}+\frac{p}{a}\frac{d\psi(a)}{da}-g_{r}\psi(a)-\frac{g_{s}}{a^{2}}\psi(a)=0, \label{fechado}
 \end{eqnarray}
 \noindent its solution is:
 \begin{eqnarray}
 \psi(a)=a^{\frac{(1-p)}{2}}\Big[ic_{1}J_{\nu}(\sqrt{-g_{r}}a)+c_{2}Y_{\nu}(\sqrt{-g_{r}}a)\Big], \label{fechado1}
 \end{eqnarray}
\noindent where $ J_{\nu} $ is the function of the first order Bessel, and $Y_{\nu} $ is the function of second-order Bessel and $\nu $ is given by:
\begin{eqnarray}
\nu=\frac{\sqrt{(p-1)^{2}+4g_{s}}}{2}. \label{C80}
\end{eqnarray}
\noindent It is possible to write the phase of the function as
\begin{eqnarray}
S=\tan^{-1}\bigg[\frac{c_{1}J_{\nu}(\sqrt{-g_{r}}a)}{c_{2}Y_{\nu}(\sqrt{-g_{r}}a)}\bigg],\label{C.26}
\end{eqnarray}
\noindent and 
\begin{eqnarray}
R=a^{(1-p)/2}\sqrt{[c_{1}J_{\nu}(\sqrt{-g_{r}}a)]^{2}+[c_{2}Y_{\nu}(\sqrt{-g_{r}}a)]^{2}}.\nonumber
\end{eqnarray}
\indent As we are focusing on the singularities at the beginning of the universe,  $a \ll 1 $ , it follows that the scale factor is small. Thus, one can use the asymptotic forms of Bessel functions, which are: 
\begin{eqnarray}
J_{\nu}(\sqrt{g_{r}}a)&\sim&\bigg(\frac{\sqrt{-g_{r}}a}{2}\bigg)^{2}, \label{C24}\\
Y_{\nu}(\sqrt{g_{r}}a)&\sim&\frac{-\Gamma(\nu)2^{\nu-1}}{(\sqrt{-g_{r}}a)^{\nu}}, \quad \nu\neq0,\label{C25}
\end{eqnarray} 
\noindent where $ \Gamma(\nu) $ is the Gamma function. Using the asymptotic approximations given by equations (\ref{C24}) and (\ref{C25}) in equation (\ref{C.26}), we obtain: 
\begin{eqnarray}
S(a\ll1) \sim-\frac{c_{1}}{2^{2\nu-1}c_{2}\Gamma(\nu)\Gamma(\nu+1)}\Big(\sqrt{-g_{r}}a\Big)^{2\nu},
\quad \textrm{to}\quad \nu\neq0. \nonumber
\end{eqnarray}
\noindent To determine the scaling factor in function of time, must be replaced in the above equation the relationship given by equation (\ref{C22}) and then integrating over time, obtaining:
\begin{eqnarray}
a(t) = \left\{ \begin{array}{ll}
\left[(3-2\nu)\gamma(\nu)(g_{r})^{-2\nu}(t+t_{0})\right]^{\frac{1}{3-2\nu}}, & \textrm{to} \quad \nu\neq0,\frac{3}{2}\\
e^{\gamma(3/2)(g_{r})^{\nu}(t-t_{0})}, & \quad \textrm{to}\quad \nu=\frac{3}{2}  \label{C500}
 \end{array}\right.
\end{eqnarray}
\noindent where $\gamma=2c_{1}\nu/2^{2\nu-1}c_{2}\Gamma(\nu)\Gamma(\nu+1)$ . In the solution above, when $\nu\neq0,3/2$, this function is finite and regular for small values of the scaling factor, this way the initial singularity is removed. For the case where $ \nu = 3/2 $, the universe will have an exponential expansion. Analyzing $ \nu = 3/2 $ in equation (\ref{C80}), we get: 
\begin{eqnarray}
 p=1\pm\sqrt{9-4g_{s}}, \nonumber
\end{eqnarray}
\noindent where we see that the ordering factor depends on the coupling constant $g_{s} $ of the model.
If $\gamma(3/2)> 0 $ , we have an expansion of the universe and for $ \gamma(3/2) <0$ , there is a contraction, which do not satisfy the evolution of the early universe.\\
\indent In order to analyze the quantum mechanism that describes the beginning of the universe is necessary to determine the quantum potential. For the case $ \nu = 3/2$, we obtain:
\begin{eqnarray}
 Q(a)= -g_{r}-\frac{g_{s}}{a^{2}}-\gamma(3/2)a^{2}. \label{C150}
\end{eqnarray}
\noindent When $ Q(a) $, the first terms of the potential cancel exactly with classic potential
$V(a)=g_{r}+g_{s}/a^{2}$. And the term  $\gamma(3/2)a^{2}$ behaves similarly to the scalar potential field or the cosmological constant in inflation, as shown in He, Gao and Cai \cite{He}. Therefore we see that the quantum effects for small scale factor values are dominant before other potential, since it requires, from the beginning, that $a(t)$ has a regular behavior, thus causing an exponential expansion.\\
\indent Now qualitatively analyze the behavior of entropy in the case of the closed universe. For this, the wave function described by (\ref{fechado1}) will be rewritten in terms of the approaches given in (\ref{C24}) and (\ref{C25}), so we obtain:
\begin{eqnarray}
 \psi(a)=a^{\frac{(1-p)}{2}}\left(-\frac{ic_{1}g_{r}a^{2}}{4}-\frac{c_{2}\Gamma(\nu)2^{\nu-1}}{(\sqrt{-g_{r}}a)^\nu}\right), \label{funcao1}
\end{eqnarray}
\noindent where $D=-c_{1}g_{r}/4$ and $B=c_{2}\Gamma(\nu)2^{\nu-1}/(\sqrt{-g_{r}})^{\nu}$. Then the wave function given by (\ref{funcao1}) can be written as:
\begin{eqnarray}
 \psi(a)=a^{\frac{(1-p)}{2}}\left(iDa^{2}-\frac{B}{a^\nu}\right). \label{funcao2}
\end{eqnarray}
\indent Now we can calculate the entropy given by equation (\ref{entropia}) in the case of the closed universe when $\nu=3/2$ is obtained that:
\begin{eqnarray}
 \mathcal{S} &=&\frac{-2B^{2}\ln B}{(p+1)}a(t)^{-(p+1)}+ \nonumber \\
 &+&\frac{B^{2}(p+2)}{(p+1)}\bigg[a(t)^{-(p+1)}\left(\ln a(t)+\frac{1}{(p+1)}\right)\bigg]+ \nonumber \\
 &+& \mathcal{S}_{0}, \label{entropiafechado}
\end{eqnarray}
\noindent the solution given in (\ref{entropiafechado}) is valid only for the case where $p<-2$. Equation  (\ref{entropiafechado}) is shown in Figure 2. It is observed that as the scale factor decreases the entropy decreases.

\begin{figure}[!htb]
\centering
\includegraphics[scale=0.5]{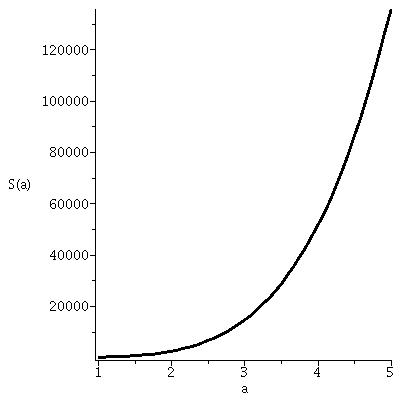}
\caption{Entangled entropy versus the scale factor, for the value of $p<-2$.}
\end{figure}

\indent Now we will express the entropy given in (\ref{entropiafechado}) in terms of time. Simply just replace the solution (\ref{C500}) for the case $\nu=3/2$ in equation (\ref{entropiafechado}), obtaining:
\begin{eqnarray}
 \mathcal{S} &=&\frac{-2B^{2}\ln B}{(p+1)}e^{-\gamma(3/2)(g_{r})^{\nu}(p+1)(t+t_{0})}+ \frac{B^{2}(p+2)}{(p+1)}*\nonumber \\
 &*&\bigg[e^{-\gamma(3/2)(g_{r})^{\nu}(p+1)(t+t_{0})}\bigg(-\gamma(3/2)(g_{r})^{\nu}(p+1)(t+t_{0})+ \nonumber \\
 &+&\frac{1}{(p+1)}\bigg)\bigg]+\mathcal{S}_{0}, \label{entropiafechado2}
\end{eqnarray}
\noindent Note that for an initial time, entropy assumes a constant value.

\subsection{\label{sec:level4}Open universe}
\qquad In this case, when the universe is opened, the Wheeler-DeWitt equation (\ref{C10}) can be written as:
 \begin{eqnarray}
 \frac{d^{2}\psi(a)}{da^{2}}+\frac{p}{a}\frac{d\psi(a)}{da}-g_{r}\psi(a)+\frac{g_{s}}{a^{2}}\psi(a)=0, \label{C789}
 \end{eqnarray}
 \noindent whose analytical solution is:
 \begin{eqnarray}
 \psi(a)=a^{\frac{(1-p)}{2}}\Big[ic_{1}J_{\mu}(\sqrt{-g_{r}}a)+c_{2}Y_{\mu}(\sqrt{-g_{r}}a)\Big], \label{C790}
 \end{eqnarray}
\noindent where $J_{\mu}$ is the function of the first order Bessel, and $Y_{\mu}$ is the function of second-order Bessel and $\mu$ is given by:
\begin{eqnarray}
\mu=\frac{\sqrt{(p-1)^{2}-4g_{s}}}{2}.
\end{eqnarray}
\noindent We can write the phase function as:
\begin{eqnarray}
S=\tan^{-1}\bigg[\frac{c_{1}J_{\mu}(\sqrt{-g_{r}}a)}{c_{2}Y_{\mu}(\sqrt{-g_{r}}a)}\bigg],\label{C26}
\end{eqnarray}
\noindent and 
\begin{eqnarray}
R=a^{(1-p)/2}\sqrt{[c_{1}J_{\mu}(\sqrt{-g_{r}}a)]^{2}+[c_{2}Y_{\mu}(\sqrt{-g_{r}}a)]^{2}}.\nonumber
\end{eqnarray}
\indent Using the same method for the case of closed universe, it can determine the scale factor as a function of time, which is given by:
\begin{eqnarray}
a(t) = \left\{ \begin{array}{ll}
\left[(3-2\mu)\gamma'(\mu)(g_{r})^{-2\mu}(t+t_{0})\right]^{\frac{1}{3-2\mu}}, & \textrm{for} \quad \mu\neq0,\frac{3}{2}\\
e^{\gamma'(3/2)(g_{r})^{\mu}(t-t_{0})}, & \quad \textrm{for}\quad \mu=\frac{3}{2} \label{C1000}
 \end{array}\right.
\end{eqnarray}
\noindent where $\gamma'=2c_{1}\mu/2^{2\nu-1}c_{2}\Gamma(\mu)\Gamma(\mu+1)$. For the case where $\mu=3/2$, is obtained:
\begin{eqnarray}
 p=1\pm\sqrt{9+4g_{s}}, \nonumber
\end{eqnarray}
\noindent where we see that the  ordering factor depends on the coupling constant $g_ {s}$ model and the scale factor will have a similar behavior to the closed universe.\\
\indent The quantum potential when the $a\rightarrow0$ to $\mu=3/2$ in this case is given by:
\begin{eqnarray}
 Q(a)= -g_{r}+\frac{g_{s}}{a^{2}}-\gamma'(3/2)a^{2}. \label{aberto}
\end{eqnarray}
\noindent Compared to the results found for the closed universe, there is only the term that involves constant $g_ {s}$ changes sign simultaneously with classic potential involving this same constant, and all the terms are still canceled the Similarly. Therefore, the term $\gamma'(3/2)a^{2}$ causes the exponential expansion.\\
\indent In this subsection was presented and discussed as quantum effects remove existing singularities in the early universe through the quantum potential. For small scale factor values, the quantum potential has dominant terms in relation to the classic potential. It was also seen that the spatial factor depends on the coupling constant $g_ {s}$, which will be decisive to characterize the evolution of the universe. However, when it is considering large values of the scale factor in the IR regime, did not realize additional contributions due to quantum effects, and thus there is no dependence on the choice of the factor order $p$ to the limit of low power \cite{Vakili}.The behavior of the entropy is analogous to the case of the closed universe.

\section{\label{sec:level5}Conclusions}
\qquad In this work, we studied the fundamental \hyphenation{sin-gu-la-ri-ty}singularity of the universe in ultraviolet phase (UV), and adopted the model proposed by
Ho\v{r}ava-Lifshitz. The solutions obtained were analyzed using the interpretation of quantum mechanics of de Broglie-Bohm, which when
interpreted result in removal of cosmology singularities.  To examine the solutions from the point of quantum view was also considered the ordering factor of momentum \hyphenation{o-pe-ra-tors}operators.\\
\indent More specifically three cases were analyzed, the first case, it is the end of predominantly classic potential before other terms, given by equation (\ref{C11}), whose solution is given by the wave function (\ref{funcaoonda}).From the analysis of this solution, the conditions have been established where $(p-1)^ {2}+4kg_{s}>0$, since $k$ and $g_{s}$ have different signals to each other and that $p\neq1\pm\sqrt{9+4kg_{s}}$. Using the interpretation of de Broglie-Bohm, was determined scale factor as a function of time, given by equation (\ref{C999}), where it can be concluded that this is a finite and regular function, but valid for only scale factor $a$ very small. Using equation (\ref{C13}) was calculated using the probability density function of time, which depends on the spatial factor $p$, and is valid for any geometry of space-time. Equation (\ref{C998}) shows that when the $a$ is very small, it has to be $\rho\rightarrow0$, removing the singularity. It was also determined as the entropy varies with the scaling factor, the equation (\ref{entropia1}), which was generated a graph represented by Figure 1, on which we can see that as the scale factor decreases, the entropy decreases. Subsequently, this was found to entropy varies over time, equation (\ref{entropia2}).\\
\indent In the other two cases, we studied regions near the singularities for spherical and hyperbolic geometries, and added a potential term to most, which contains the coupling constant $g_{r}$. The equation used to describe the closed universe, in this case, is given by equation (\ref{fechado}), which solution is given by (\ref{fechado1}). Using the interpretation of de Broglie-Bohm be found to equation (\ref{C500}), which describes how the scale factor evolves over time. When it is considered $\nu\neq0,3/2$, the expansion is described by a finite and regular function, valid for small values of the scale factor. On the other hand, if $\nu=3/2$, the expansion is exponential and is caused by the term $\gamma(3/2)a^{2}$ present in the quantum potential equation (\ref{C150}). Therefore, the singularity is removed due to quantum effects. Then we calculated the entropy due to the scale factor, which produces a restriction in the amount of $p$, which in this case must satisfy $p<-2$. So, a numerical analysis of these results was taken, which is shown in Figure 2. It was also possible to determine the entropy function scale factor.\\
\indent In the case that the universe is open, the equation that describes it is given by (\ref{C789}), whose wave function is expressed by equation (\ref{C790}). The results for the scale factor as a function of time, equation (\ref{C1000}) and the quantum potential, equation (\ref{aberto}), are similar to results found for the closed universe only the term it involves coupling constant $g_{s}$ changes sign simultaneously with classic potential involving the same constant. Thus, the singularity is avoided. For cases geometries spherical and hyperbolic, spatial factor $p$ depends on the coupling constant $g_{s}$. The results for the entropy are analogous to those found for the closed universe. It is emphasized in the low power limit, when the scale factor is large, quantum effects are not relevant, so no dependence on the choice of the orderning factor \cite{Horava}.\\
\indent In summary, we found a result that suggested a intimate relationship between the factor ordering and coupling constants  $g_{s}$ original of HL gravity, showing their quantum nature , finally  were considered the solutions  to  obtain the entanglement entropy that have exactly the following behavior: a increase of entanglement entropy in the limit of small scale factor with time.

\bibliographystyle{abbrv}
\bibliography{refs}

\end{document}